\documentclass[12pt]{article}

\usepackage{scicite}

\usepackage{times}

\usepackage{amsmath}
\usepackage{amsfonts}
\usepackage{amssymb}
\usepackage{graphicx}
\newcommand{\sr}{$^{87}\mathrm{Sr}$}
\newcommand{\lmt}{$10^4 \hbar k$}
\newcommand{\intercomb}{$^1S_0-^3P_1$}

\newenvironment{sciabstract}{%
\begin{quote} \bf}
{\end{quote}}

\title{Circulating pulse cavity enhancement as a method for extreme momentum transfer atom interferometry}

\author
{Rustin Nourshargh${}^{1\ast}$, Samuel Lellouch${}^{1\ast}$, Sam Hedges${}^{1}$, Mehdi Langlois${}^{1}$,\\ Kai Bongs${}^{1}$, Michael Holynski${}^{1\dagger}$\\
\\
\normalsize{${}^{1}$School of Physics and Astronomy, University of Birmingham}\\
\normalsize{University of Birmingham, Edgbaston, B15 2TT, UK}\\
\\
\normalsize{$^\ast$These authors contributed equally to this work.}\\
\normalsize{$^\dagger$To whom correspondence should be addressed; E-mail: M.Holynski@bham.ac.uk}
}

\date{}

\begin{document} 


\baselineskip24pt

\maketitle

\begin{sciabstract}
Large scale atom interferometers promise unrivaled strain sensitivity to mid-band (0.1 - 10~Hz) gravitational waves, and will probe a new parameter space in the search for ultra-light scalar dark matter. These atom interferometers require a momentum separation above \lmt{} between interferometer arms in order to reach the target sensitivity. Prohibitively high optical intensity and wavefront flatness requirements have thus far limited the maximum achievable momentum splitting. We propose a scheme for optical cavity enhanced atom interferometry, using circulating, spatially resolved pulses, and intracavity frequency modulation to overcome these limitations and reach \lmt{} momentum separation. We present parameters suitable for the experimental realization of \lmt{} splitting in a 1~km interferometer using the 698~nm clock transition in \sr{}, and describe performance enhancements in 10~m scale devices operating on the 689~nm intercombination line in \sr{}. Although technically challenging to implement, the laser and cloud requirements are within the reach of upcoming cold-atom based interferometers. Our scheme satisfies the most challenging requirements of these sensors and paves the way for the next generation of high sensitivity, large momentum transfer atom interferometers.
\end{sciabstract}

\section*{\label{sec:Intro}Introduction}

Atom interferometers (AI) are high-precision devices with a wide range of potential applications~\cite{Cronin2009,McGuirk2002}, from gravitational measurements~\cite{Rosi2014,Fixler2007} to inertial sensing, navigation~\cite{Durfee2006,Peters1999}, and metrology~\cite{Weiss1994,Cadoret2008,Schlippert2014}. The measurement is inferred from the interference between two atomic clouds which are split, reflected, and recombined through coherent pulses of light~\cite{Kasevich1991,Giltner1995}. Sensitivity is enhanced by increasing the space-time area of the interferometers~\cite{McGuirk2000} through large momentum transfer (LMT) techniques, and by increasing the total interrogation time. Recent proposals~\cite{graham2017mid,canuel2019elgar,Badurina2020} use AI for gravitational wave~\cite{Dimopoulos2008,Chaibi2016,Hild2011,Sathyaprakash2012,AdvancedLIGO2015,Acernese2014} and dark-matter~\cite{Arvanitaki2018,ElNeaj2020} detection, as they enable exploration of currently inaccessible frequency bands. These proposals require sensors with kilometer scale baselines and momentum separations of order \lmt.

Traditional atom interferometers rely on multiphoton interactions: two-photon Bragg~\cite{Chiow2011} or Raman~\cite{McGuirk2000} transitions, or multiphoton Bragg~\cite{Muller2008} transitions. Such techniques are sensitive to differential laser phase noise, making them challenging to implement in very large baseline systems where propagation delay causes the laser noise to dominate the overall noise budget. Differential single photon interferometry benefits from greatly reduced susceptibility to laser noise and is the most promising technique for future large-scale interferometers~\cite{Graham_2013, graham2016resonant}. Large momentum transfer with sequential single-photon transitions has been demonstrated in  $^{88}\mathrm{Sr}$~\cite{Abend2016,McDonald2013,Mazzoni2015}, achieving interferometry with a momentum separation of $141\hbar k$~\cite{Rudolph2020LMT689nm}. Extending the technique to \lmt{} raises significant challenges. Spontaneous emission must be minimized during the interferometry sequence, motivating the use of transitions to long lived excited states such as those used in atomic clocks~\cite{Ushijima2015,Bloom2014}. The weak coupling to these long lived excited states requires high optical powers to match the Rabi frequencies that can be achieved on short lived states.

The 698~nm optical clock transition in \sr{} is a suitable candidate for single photon interferometry. With an  excited state lifetime of 150s ($\Gamma=2\pi \cdot 1~\mathrm{mHz}$), spontaneous emission is a negligible source of decoherence~\cite{Santra2004,Hu2020}. With microsecond $\pi$-pulses and therefore MHz Rabi frequencies, the \lmt{} pulse sequence is completed in a small fraction of the total interrogation time. The Rabi frequency is $\Omega=\Gamma\sqrt{I/2I_S}$, with $I_S=0.4~\mathrm{pW/cm}^{2}$, so microsecond pulses require prohibitively large intensities of several $\mathrm{kW/cm}^{2}$. Delivering these pulses with uniform phase also demands flat optical wavefronts~\cite{Trimeche2017Wavefront}. These constraints have prevented momentum separations of \lmt{} from being achieved thus far.

Optical cavities offer a potential solution to both of these problems~\cite{Ye2003}: resonant power enhancement reduces the required input power, and spatial filtering improves wavefront flatness. Impressive early demonstrations of cavity enhanced interferometry have shown the potential of these systems~\cite{Hamilton2015Cavity,Riou_2017}. However, cavity power enhancement is only effective if the pulse duration exceeds the cavity lifetime. For pulses shorter than, or comparable to the cavity lifetime the cavity fails to reach its maximum power enhancement. The intracavity field will persist long after the input light is removed resulting in significant pulse elongation. This bandwidth limit described by Alvarez et al.~\cite{Alvarez17BWLimit} restricts the maximum length-finesse product that can be achieved without significant pulse elongation. In a kilometer scale cavity, the bandwidth limit restricts the maximum finesse to $\mathcal{F}\leq10$, significantly reducing the mode filtering and power enhancement benefits.

We present a scheme based on circulating, spatially resolved pulses to overcome the bandwidth limit and satisfy the challenging pulse requirements for \lmt{} LMT. A circulating pulse drives a $\pi$-pulse once per round trip, removing the need to couple a new pulse into the cavity for each beam splitter. This avoids lifetime elongation and overcomes the bandwidth limit, dramatically increasing the possible lengths and cavity finesses that can be exploited. We present parameters for a cavity and laser system capable of \lmt{} momentum separation on the 698~nm transition in \sr{} in only 200~ms.

\section*{Results}

\begin{figure}[ht]
	\centering
	\includegraphics[width=.8\columnwidth ]{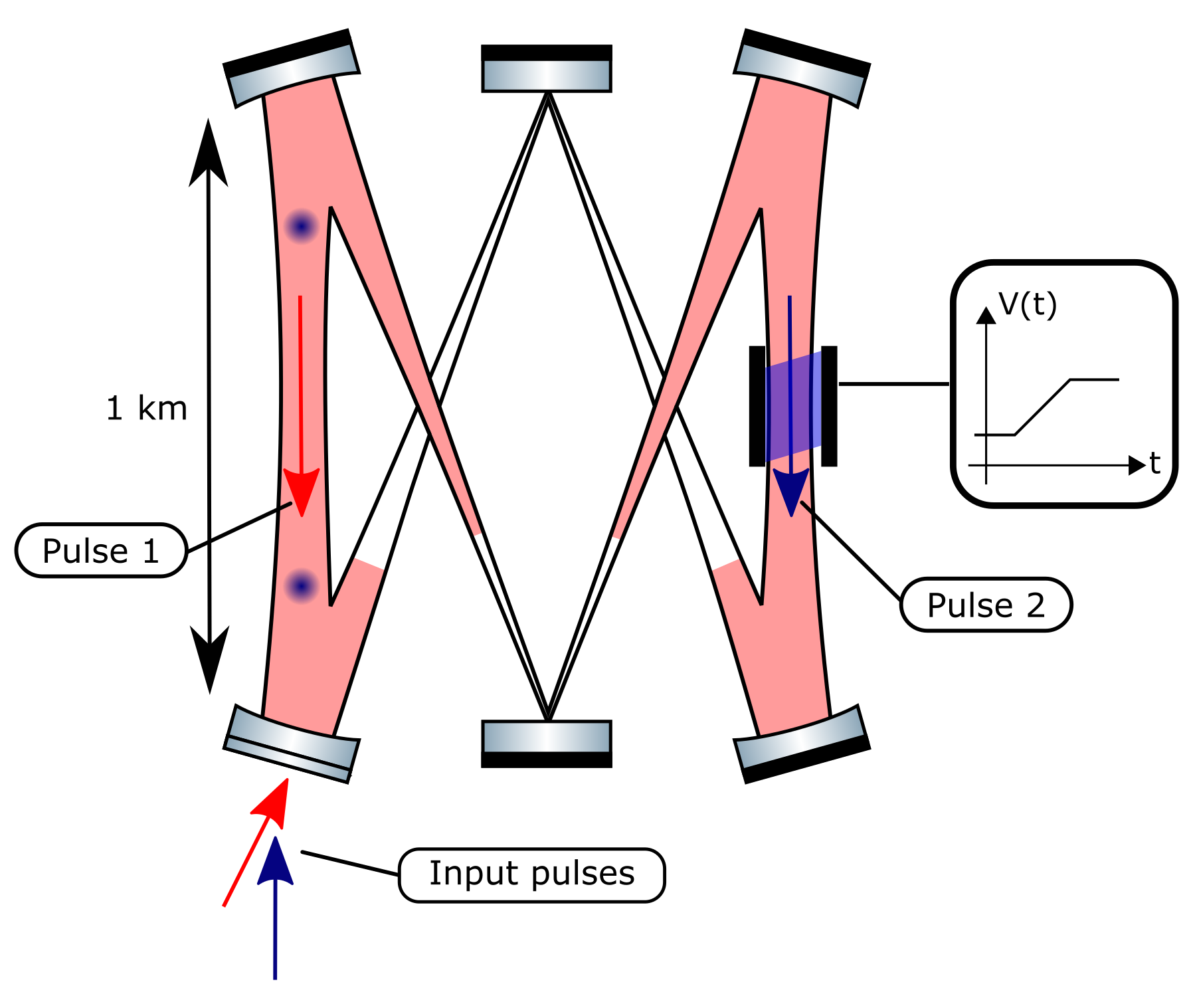}
	\caption{Circulating pulses in each of the two running wave modes in this 6~km round trip cavity. On each round trip, additional light is coupled into the cavity to compensate for losses. Serrodyne modulation, applied through a Pockels cell, shifts the frequency of each pulse on each pass to compensate for Doppler shifts. The pulse durations are maximized within the constraint that only one pulse may pass through the atoms (blue) or Pockels cell at a time.
	}
	\label{fig:CavitySchematic}
\end{figure}

\begin{figure}[h!]
	\centering
	\includegraphics[width=1.\columnwidth ]{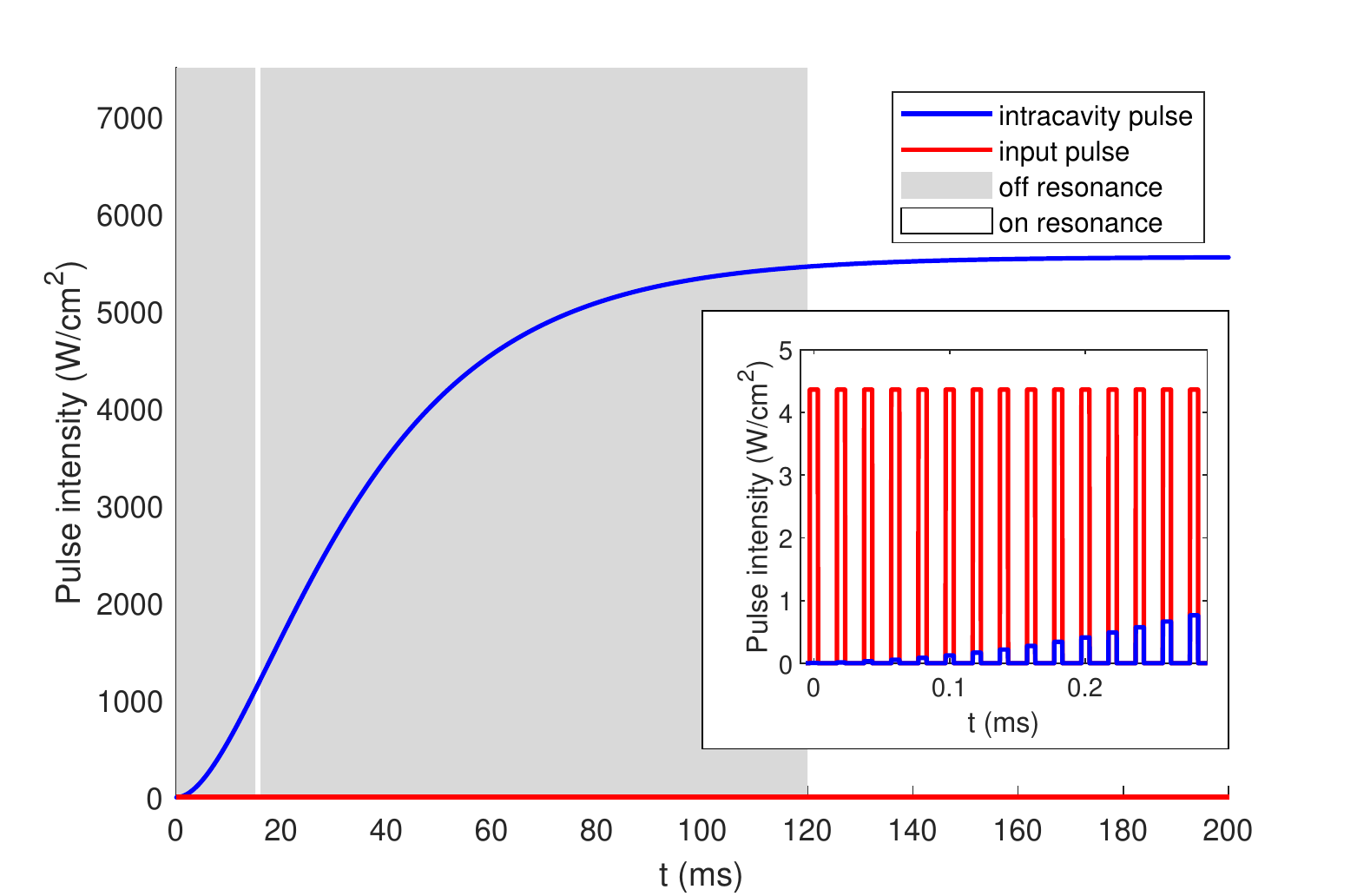}
	\caption{Cavity response to a periodic train of spatially-resolved coherent square pulses as a function of time. The input intensity $I_{in}$ is plotted in red and the intracavity intensity $I_{circ}$ in blue. The main plot shows the only the intensity maxima, with the full time-dependence of the first input pulses shown in the inset. The system parameters match those described in the text, $L=6~\mathrm{km}$, $\mathcal{F}=4000$, $\tau=6~\mu \mathrm{s}$. The intracavity intensity increases as input pulses coherently add, before saturating at its stationary cavity-enhanced value, which is exactly that of a $\pi$-pulse. Grey-shaded areas indicate when the light is off-resonant with the atomic transition. It is shifted on-resonance for one round trip at $t=15~\mathrm{ms}$ to drive the initial $\pi/2$-pulse, and again after $120~\mathrm{ms}$ to perform the LMT sequence; see Fig.~\ref{fig:ClockAI}.
	}
	\label{fig:BuildUp}
\end{figure}

Consider a traveling wave cavity into which we periodically couple short, spatially-resolved pulses of light. The cavity has a 1~km baseline and a round trip path length of 6~km (see Fig.~\ref{fig:CavitySchematic}). We use a pulse duration $\tau = 6~\mu \mathrm{s}$, which corresponds to a physical length of 1.8~km and satisfies the condition for spatially resolved pulses; see Materials and Methods. Successive pulses coupled into a single circulating mode are separated by the cavity round-trip time ($20~\mu \mathrm{s}$) such that they constructively interfere. This periodic train of pulses forms a comb in the frequency domain, but in this regard it is identical to free space LMT schemes using periodic short pulses. Light is coupled in away from atomic resonance, and shifted onto resonance using Serrodyne modulation once a stationary regime has been reached (see Fig.~\ref{fig:BuildUp}), resulting in a high-intensity pulse circulating inside the cavity. The circulating pulse intensity is adjusted to drive a $\pi$-pulse on the atoms loaded in the cavity, delivering a momentum kick of $\hbar k$ on each round trip. To ensure successive momentum contributions add constructively, we alternate the pulse direction by populating both circulating modes of the cavity. After 100~ms, $5000$ round trips have taken place and the target splitting of \lmt{} is achieved.

In this beam splitter sequence, momentum is only imparted to one arm of the interferometer (referred to as the \textit{fast-arm}), leaving the other (\textit{slow-arm}) unaffected. Lifting the initial degeneracy between arms requires careful treatment, outlined in Materials and Methods.

\paragraph{Cavity build-up and pulse requirements}
 In the following analysis we consider only one of the two running wave modes of the cavity containing the circulating pulses, but note that the results apply equally to both. The input intensity is fixed by the requirement that the circulating pulse drives a $\pi-$transition on the atoms. The response of the intracavity field to the train of phase coherent input pulses is obtained by solving the propagation equation where $t$ is the transmission coefficient of the input mirror, $(1-\gamma)$ is the total round trip loss, and $i= \sqrt{-1}$. If we select $t=\gamma / \sqrt{2}$ such that the cavity is impedance matched, this is given by
\begin{equation}
    \tilde{E}_{circ}(t)=it \tilde{E}_{in}(t)+\gamma \tilde{E}_{circ}(t-\tau_{rt})
    \label{eq:PropEq}
\end{equation}

The circulating pulse amplitude increases with each successive input pulse before reaching a stationary value where further pulses only compensate for round-trip losses; see Fig.~\ref{fig:BuildUp}. The stationary pulse intensity is cavity enhanced $I_{circ}(t)=(\mathcal{F}/\pi)I_{in}(t)$ where $\mathcal{F}$ is the finesse~\cite{siegman1986lasers}. Cavity enhancement allows this scheme to achieve $\pi-$pulses inside the cavity with reduced input laser power. A finesse of 4000 enables $6~\mu \mathrm{s}$ $\pi-$pulses with an input intensity, at the center of the beam, of only $4.4~\mathrm{W}/\mathrm{cm}^{2}$, compared to $5.6~\mathrm{kW}/\mathrm{cm}^{2}$ without cavity enhancement. This represents a 1000-fold reduction in required laser power, putting it within reach of existing laser technology, see materials and methods.

The total power required depends on the radius of the laser beams, with lower bounds of both atomic and optical origin. High-fidelity pulses can only be achieved if the intensity profile does not vary across the atomic cloud. With a beam to cloud diameter ratio of 20, residual intensity variations have a negligible contribution to fidelity loss. A typical cloud of radius of $200~\mu \mathrm{m}$ would require a $1/e^2$ beam radius of 4~mm~\cite{Nicholson2015}, and a total input power of 1.1~W ~\cite{Rudolph2020LMT689nm,Hu2020}. However, divergence of the Gaussian beam over the 1~km arm length is the more stringent constraint. The minimum beam radius at 500~m, $w=1.5~\mathrm{cm}$, is produced with a beam waist $w_0 = 1.05~\mathrm{cm}$. This results in a minimum required input laser power of 16~W. Since the beam radius, and therefore intensity, varies with position along the optic axis, longer atomic freefall times may necessitate increased beam diameters to achieve required axial intensity homogeneity.

\begin{figure}[h!]
	\centering
	\includegraphics[width=0.9\columnwidth ]{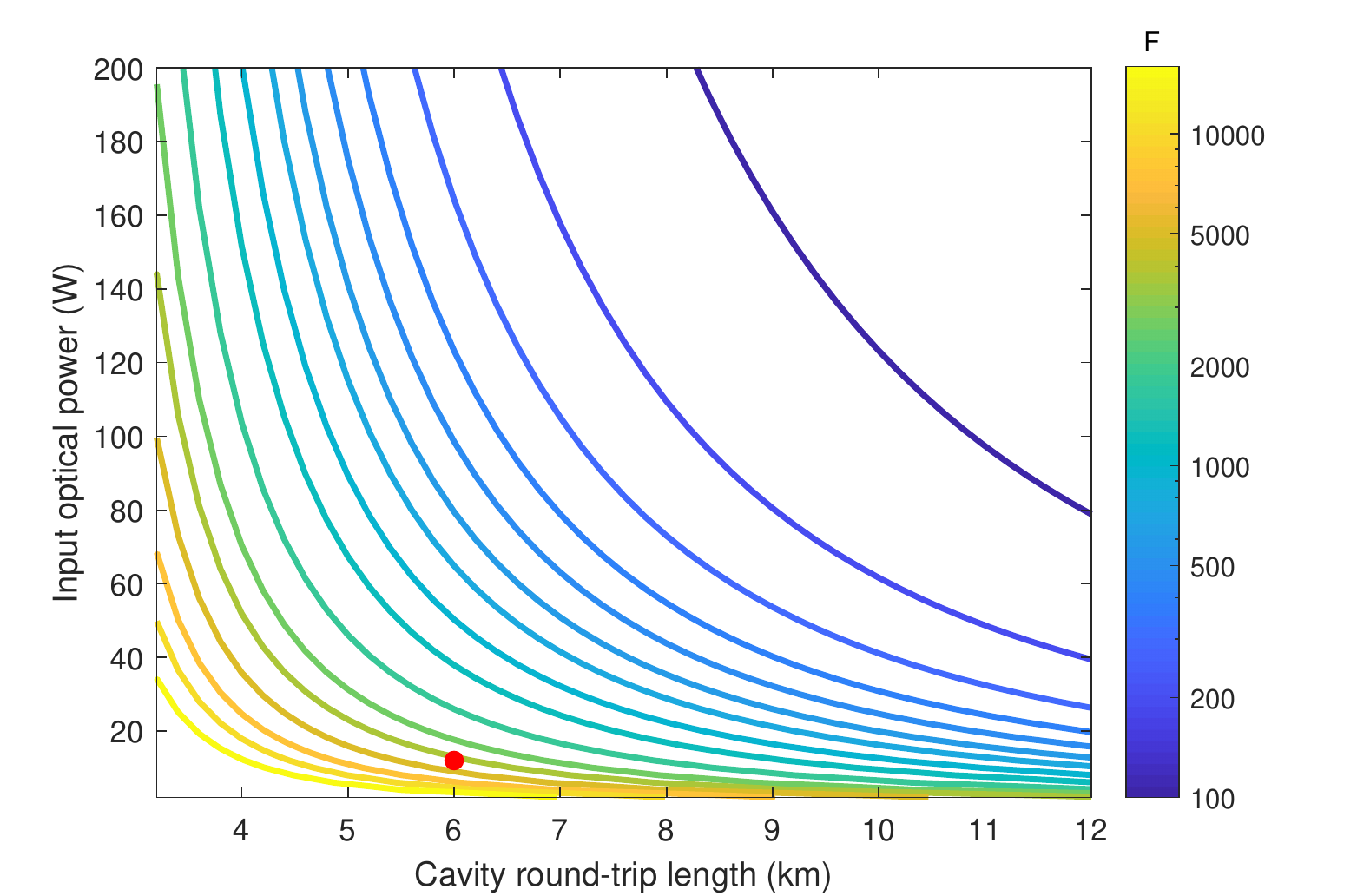}
	\caption{Required input optical power as a function of cavity round-trip length and finesse. The cavity baseline is fixed at 1~km, and the maximum pulse duration, minimizing power requirements, is selected. This is the limiting value for the pulses to avoid spatial overlap at the atoms (see Materials and Methods). The parameters used throughout are indicated by the red dot.
	}
	\label{fig:Scalings}
\end{figure}

The required input optical power decreases linearly with finesse and quadratically with pulse duration. The need to avoid pulse overlap at the atoms constrains the maximum pulse duration; see Materials and Methods. Fig.~\ref{fig:Scalings} shows the minimum optical input power required as a function of round-trip path length, for cavities with a baseline of 1~km and differing values of finesse. The clear reduction in input power motivates the use of large high-finesse cavities. The increase in the time taken to reach the stationary regime (linear in both length and finesse) is not a limiting factor in the proposed scheme.

\paragraph{Interferometric sequence}
Pulses of light are coupled into the cavity far-detuned from atomic resonance, avoiding unintentional interactions during the build-up phase. Once the target amplitude has been reached (white areas in Fig.~\ref{fig:BuildUp}), intracavity serrodyne modulation is used to shift the circulating pulse onto atomic resonance. Serrodyne modulation is generated by applying a linear ramp to an intracavity Pockels cell. High modulation efficiencies and low losses are essential to reaching the required pulse amplitude; some experimental steps towards realizing this are discussed in the materials and methods.

\begin{figure}[h!]
	\centering
	\includegraphics[width=.8\columnwidth]{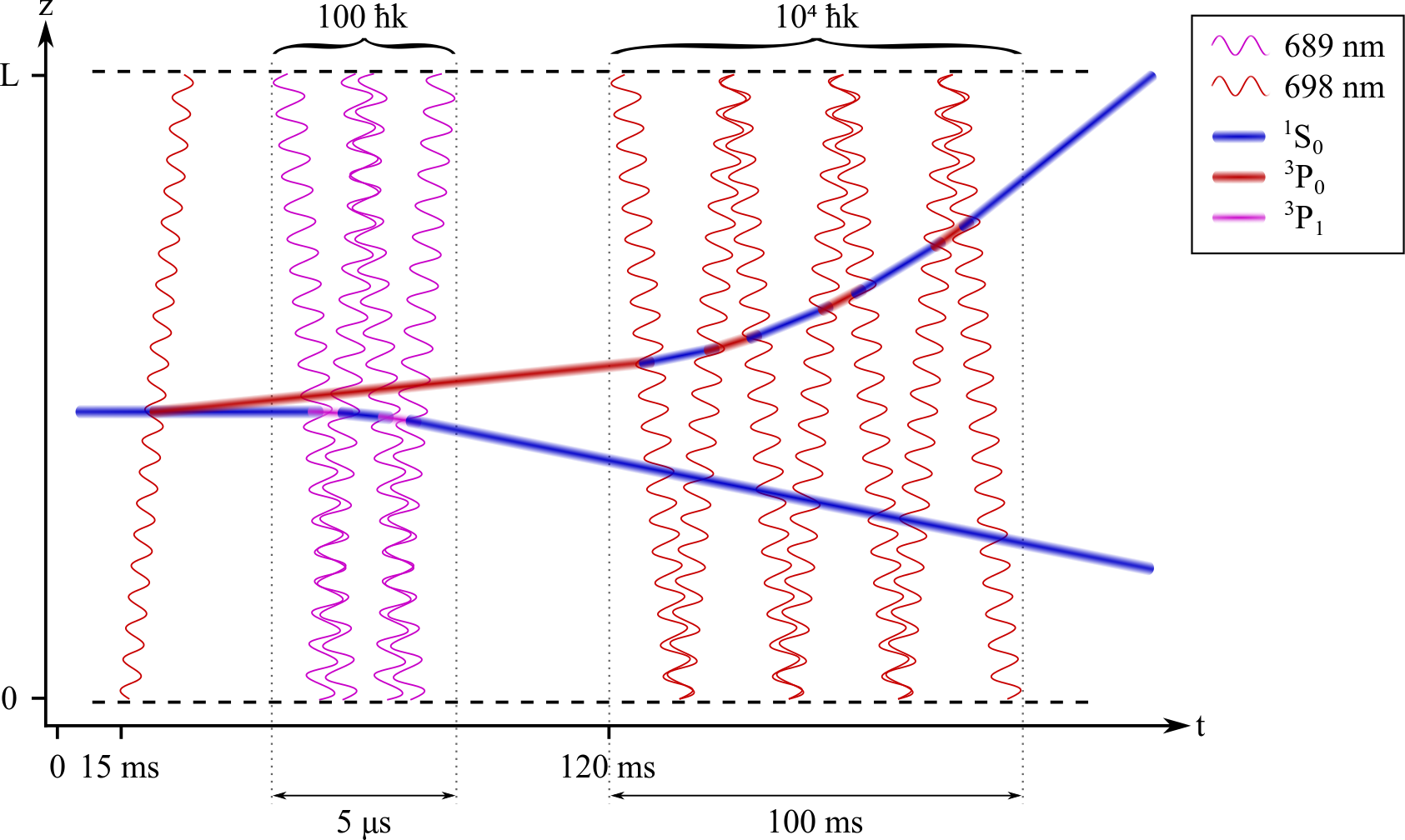}
	\caption{Space-time diagram for the large momentum transfer atom interferometry scheme. An initial 698~nm (red, wavy) $\pi/2$-pulse transfers the atoms into a superposition between the ground state, slow-arm (blue), and the excited state, fast-arm (red). Sequential counterpropagating $\pi$-pulses on the 689~nm transition (pink, wavy) deliver 100$\hbar k$ of momentum to the slow-arm only. With the frequency degeneracy lifted, the main LMT sequence begins at 120~ms, delivering \lmt{} of momentum to the fast-arm through counterpropagating $\pi$-pulses on the 698~nm transition.
	}
	\label{fig:ClockAI}
\end{figure}

At $t = 15~\mathrm{ms}$ the circulating amplitude has reached that of a $\pi/2$-pulse and one of the pulses is shifted onto resonance for a single round trip, see white vertical line on Fig.~\ref{fig:BuildUp}. This delivers a $\pi/2$-pulse and is followed by a further serrodyne shift to move the pulse away from resonance whilst the amplitude is increased further.

We propose two techniques to lift the initial arm degeneracy: imparting momentum to the slow-arm of the interferometer with the 689~nm transition, and bichromatic beam splitter pulses to track the recoil of both arms.

At $t = 120~\mathrm{ms}$, the circulating intensity has reached the value required for a $\pi$-pulse and the initial degeneracy has been lifted. Intracavity serrodyne modulation is used to shift both circulating pulses onto resonance with the fast-arm of the interferometer. On each round trip they deliver a pair of sequential counterpropagating $\pi$-pulses, imparting $2\hbar k$ of momentum. Recoil shifts and gravitationally induced Doppler shifts are compensated every round trip with further serrodyne modulation. After $5\times 10^3$ round trips, a momentum separation of \lmt{} between the interferometer arms has been achieved. The circulating pulses are either dumped from the cavity or serrodyne shifted back away from resonance in preparation for the next beam splitter. The total sequence duration is less than 1~s, a duration over which spontaneous emission losses remain negligible. A similarly constructed beam splitter sequence is used to close the interferometer.

\paragraph{Doppler shift and phase compensation}
In addition to its primary role of shifting pulses on and off the atomic resonance, serrodyne modulation is used to compensate for Doppler shifts in the sequence. Fig.~\ref{fig:ClockAI} shows the fast-arm recoiling upwards on each successive beam splitter pulse. The Doppler effect will cause the downward-going pulse to appear blueshifted, whilst the upward-going pulse appears redshifted. We compensate for this by applying negative serrodyne shifts to the downward-going pulse and positive shifts to the upward-going pulse. The shifts must be applied to the two pulses separately, requiring them to pass through the Pockels cell at different times. 

Serrodyne modulation changes the phase relationship between the circulating and input light. To ensure constructive interference continues, it is vital that the circulating and input pulses remain in phase. Phase compensation can be achieved through active control of the input light or additional voltages on the Pockels cell; see Materials and Methods.  

\begin{figure}[h!]
	\centering
	\includegraphics[width=1.\columnwidth ]{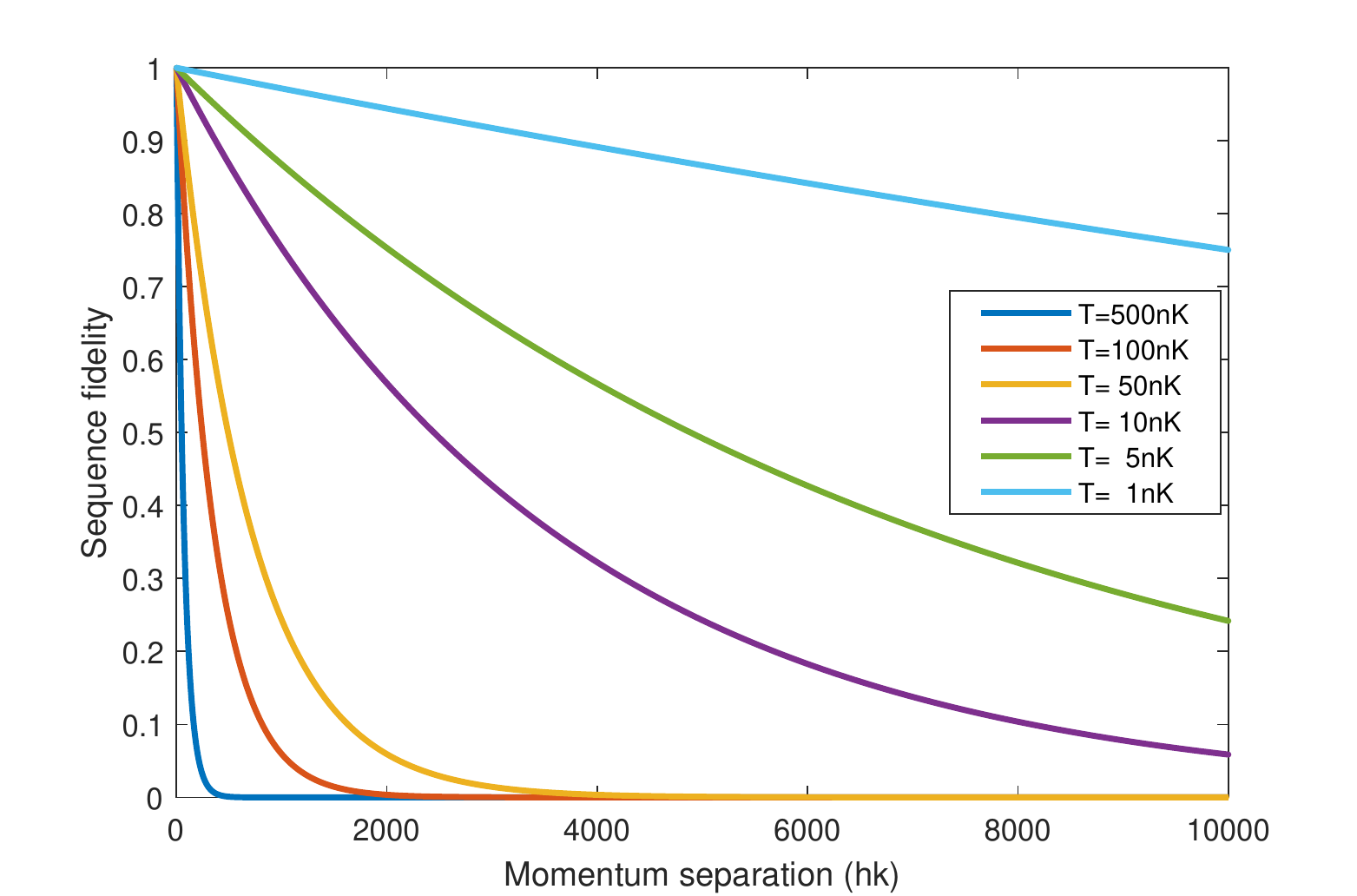}
	\caption{Fidelity of the whole LMT sequence as a function of momentum separation, for various atomic temperatures. The sequence fidelity is obtained as the cumulative product of individual pulse fidelities, which are obtained by averaging the transition probability over the cloud instantaneous spatial and velocity distributions. The initial cloud radius is $200~\mu \mathrm{m}$ and the beam radius is 1.5~cm.  To obtain a sufficient contrast at the end of the \lmt{} LMT sequence, a cloud temperature less than 10~nK is required. Cloud expansion, Doppler shift compensation and spontaneous emission are included in the model. The pulse bandwidth is neglected, but any resulting errors can be compensated with frequency and intensity adjustments~\cite{Akerman_2015,Nadlinger2016}.
	}
	\label{fig:Fidelity}
\end{figure}

\paragraph{High-fidelity LMT}
Reaching a momentum transfer of \lmt{} requires $10^4$ beam splitter pulses, which must be of high-fidelity to maintain contrast through the interferometer. There are several factors limiting the fidelity of an atom-optic pulse: the thermal velocity spread of the cloud (which results in a dispersion in the Doppler shifts), the spatial variations of the beam intensity at the scale of the cloud, and the spontaneous emission losses. Although this discussion is not specific to our scheme, we demonstrate that we can maintain sequence fidelity at \lmt{}, and identify the limiting parameter. The fidelity improves with increased beam-to-cloud diameter ratio, with a strong dependence up to a ratio of 20. At this ratio the intensity is approximately constant at the scale of the cloud and further increasing the beam diameter results in only a minor improvement in fidelity. The fidelity also improves for reduced atom cloud temperature, and in our scheme temperature is the limiting parameter. This arises from the Doppler shift dispersion inside the cloud, and not from the spatial expansion of the cloud during its propagation, which remains negligible. For a cloud radius of $200~\mu \mathrm{m}$ (beam-to-cloud ratio of 75), the sequence fidelity is plotted for various cloud temperatures, see Fig.~\ref{fig:Fidelity}. We observe that to achieve sufficient contrast at the output of the interferometer, a temperature of a few nanokelvin is required for \lmt{} LMT~\cite{Campbell90}. 

\paragraph{Implementation in smaller cavities}
This scheme could be tested on a smaller system operating on the 689~nm \intercomb{} transition in \sr{}, where higher Rabi frequencies enable shorter pulses and correspondingly smaller cavities. A 40~m-round-trip cavity of finesse 1000, with 16~ns input pulses every 133~ns, will require only 1.2~W of laser power to implement $\pi$-pulses with a beam waist of $w_0 = 15~\mathrm{mm}$. Due to the large pulse bandwidth, no Doppler shift compensation is required as both arms are addressed with high-fidelity. Whilst spontaneous emission will limit the performance on this transition below the target of \lmt{} momentum separation, it will still provide a useful performance enhancement to these lab-scale systems, and serve to validate the scheme. 

\section*{Discussion}

We have presented a novel scheme for cavity enhanced atom interferometry to enable extremely large momentum transfer beam splitters in large scale atom interferometers. Intracavity serrodyne modulation allows circulating, spatially resolved pulses to be recycled, overcoming cavity lifetime elongation and the pulse bandwidth limit. Serrodyne modulation is used to shift the pulse on and off resonance and compensate for photon recoils and gravitational Doppler shifts. 

We analyze the case of a kilometer scale atom interferometer for gravitational wave detection, and find that a cavity with 6~km round trip path length and a finesse of $\mathcal{F}= 4000$ can generate a circulating pulse intensity of $5.6~\mathrm{kW}/\mathrm{cm}^{2}$ with only $4.4~\mathrm{W}/\mathrm{cm}^{2}$ at the input. When applied to the \sr{} 698~nm clock transition this enables 6~$\mathrm{\mu s}$ $\pi$-pulses in a mode of 1.5~cm radius with an input laser power of only 16~W. The overall sequence fidelity is found to be limited by the temperature of the atomic cloud. A beam splitter with \lmt{} momentum separation between the arms and combined fidelity $>0.25$ requires a vertical velocity spread of 1.2~mm/s, corresponding to a temperature selectivity of $5~\mathrm{nK}$. Further reductions in cloud temperature increase fidelity and hence the maximum total momentum separation. This is the first practical, albeit challenging, approach to the generation of short, high-fidelity pulses on this transition.

Circulating pulse interferometry can also be applied to 10~m atom interferometers operating on the intercombination line in \sr{} at 689~nm. Resonant power enhancement and spatial mode filtering will allow for an increase in fidelity and therefore the possible momentum separation in these systems, improving sensitivity.

By overcoming the bandwidth limit we have enabled optical cavity enhancement of the largest proposed atom interferometers. Circulating pulse cavity enhanced atom interferometry offers a step change in large momentum transfer atom interferometry. With a thousand-fold increase in optical intensity, it enables large scale atom interferometers for gravitational wave detection and new tests of fundamental physics.

\section*{Materials and Methods}

\paragraph{Initial arm degeneracy}
Our scheme relies on velocity selective pulses such that only one arm of the interferometer is addressed, leaving the other unaffected. However, the large Rabi frequency ($\simeq 80~\mathrm{kHz}$) and small initial frequency separation from the first $\pi/2$-pulse (9.4~kHz) results in the two arms being initially degenerate. The two arms cannot be discriminated in frequency with high-fidelity until the frequency separation exceeds the pulse bandwidth. Since one recoil of $\hbar k$ produces a Doppler shift of $\hbar k^2/m=2\pi \cdot 9.4~\mathrm{kHz}$, $\simeq 10-100$ beam splitters are required to lift this degeneracy. 

One method to open the interferometer is to create the initial momentum separation with $\pi$-pulses on the $^1S_0-^3P_1$ 689~nm transition. The initial $\pi/2$-pulse on 698~nm splits the atoms into a superposition of the long lived excited state and the ground state, referred to as the fast and slow-arm respectively. We apply one hundred sequentially counterpropagating $\pi$-pulses on the 689nm transition, imparting $100\hbar k$ of momentum to the slow-arm on the interferometer whilst leaving the fast-arm in the excited state unaffected. The resulting Doppler shift is $100\hbar k^2/m \simeq 2\pi\cdot 1 ~\mathrm{MHz}$, greatly exceeding the Rabi frequency of the clock pulses, ensuring that the fast-arm of the interferometer can be uniquely addressed. 

The 689nm transition allows for much higher Rabi frequencies $\sim 100~\mathrm{MHz}$, enabling 10~ns $\pi$-pulses without cavity enhancement. The 100 pulse sequence is completed within $5~\mu \mathrm{s}$: $3.3~\mu \mathrm{s}$ for the pulses to propagate 1~km between the clouds and $\simeq 2~\mu \mathrm{s}$ for 100 pulses. The pulse bandwidth is large compared to the total Doppler shift so we achieve high-fidelity beam splitters without Doppler compensation~\cite{Rudolph2020LMT689nm}. The $^3P_1$ state has a lifetime of $21.6~\mu\mathrm{s}$ requiring a rapid pulse sequence ending in the ground state to minimise losses due to spontaneous emission. This beam splitter sequence can be accommodated within the general scheme depicted on Fig.~\ref{fig:BuildUp} and performed during the cavity build-up time, when the 698~nm pulses are far detuned from atomic resonance.

An alternative approach circumvents the issue of arm degeneracy, using only the 698nm clock transition, by utilizing two polarization modes in each pulse. The two polarization modes are independently serrodyne shifted to track the Doppler shifts experienced in both arms of the interferometer, redshifting one polarization whilst blueshifting the other. As both interferometer arms are now recoiling, the total number of beam splitter pulses required to achieve a given momentum separation is halved. This scheme requires two Pockels cells to apply a phase shift to one of the two polarizations whilst leaving the other unchanged. Cells based on $\bar{4}2m$ class crystals have this property if the $r_{63}$ coefficient is employed in a transverse configuration~\cite{Yariv1975}. This geometry results in a residual birefringence from each crystal, which we compensate by arranging the two crystals orthogonally. Phase compensation can be achieved with the same methods discussed in the results, but must now be applied to both polarizations simultaneously.

This dual polarization, dual frequency scheme requires careful control of the light intensities on a pulse-by-pulse basis to ensure this bichromatic light field delivers high-fidelity $\pi$-pulses to both arms on each round trip. The details of the specific powers and frequencies required to achieve this are beyond the scope of this paper.

\paragraph{Maximum pulse duration}
The scaling of input power with pulse duration motivates the use of the longest pulses possible. However, the interferometry scheme described in this work requires successive pulses to alternate in direction and to be temporally distinct when they pass through the atoms and the Pockels cell. This constrains the maximum pulse duration.

\begin{figure}[h]
	\centering
	\includegraphics[width=1.\columnwidth ]{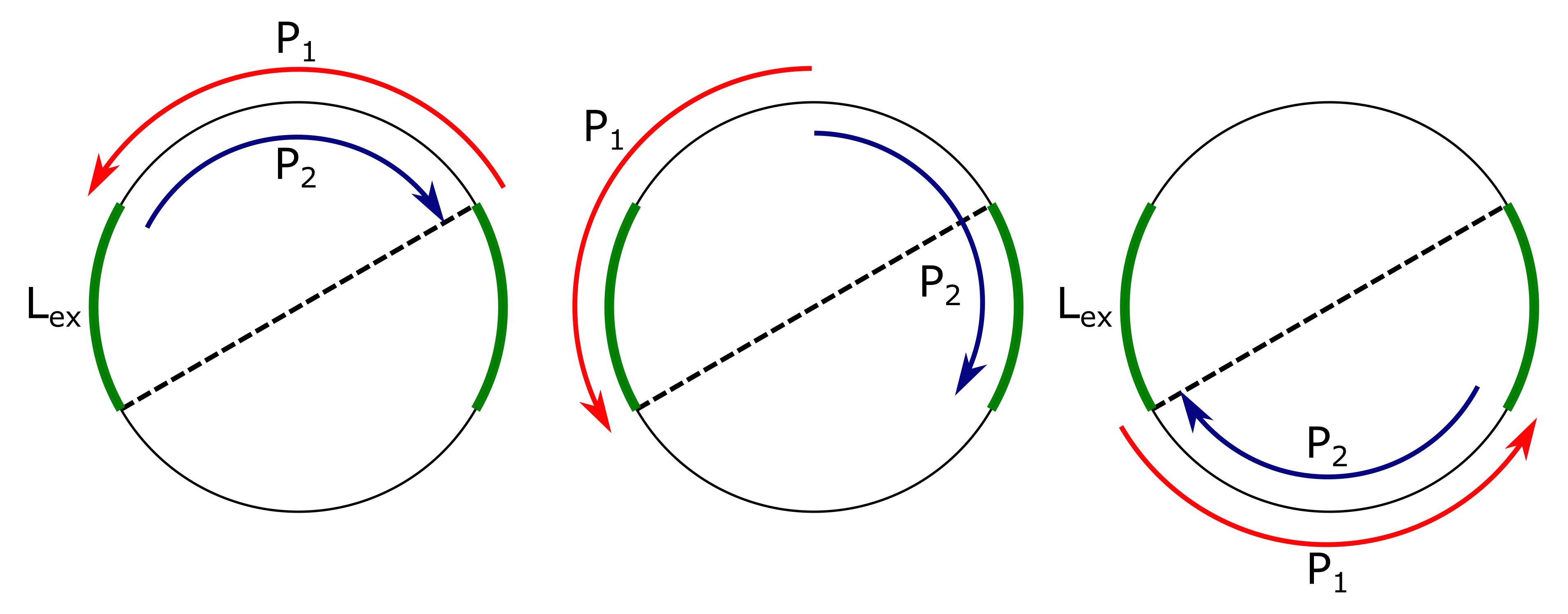} 
	\caption{This figure shows counterpropagating pulses of duration $\tau_{max}$ circulating round the cavity represented by the circumference of the circle. In the first frame pulse-2 has just left the exclusive length and pulse-1 is about to enter it. The relationship between the pulse duration, exclusive length and round trip path length is clear from the first and third frames: $L_{RT}/2 = L_{ex} + c\tau_{max} $
	}
	\label{fig:PulseDurationLimit}
\end{figure}

We define an exclusive length $L_{ex}$ as the region within the cavity where the pulses must not overlap. Two pulses, of equal limiting duration $\tau_{max}$, propagating in opposite directions, will fully overlap when one pulse leaves the exclusive length and the other is about to enter. This condition is repeated on both sides of the exclusive length. Symmetry considerations show that there is an equal exclusive length on the opposite side of the cavity. Fig.~\ref{fig:PulseDurationLimit} illustrates the locations of the pulses and exclusive length in a generalized ring cavity. Summing the exclusive length and pulse length the condition becomes clear:

\begin{equation}
\tau_{max} = \frac{1}{2c}\left(L_{RT} -2L_{ex}  \right) 
\end{equation}

For pulses with $\tau \leq \tau_{max}$ and correct input timings, singular occupation of the exclusive lengths will be achieved. Pulse durations approaching $\tau_{max}$ reduce performance requirements for the cavity and input lasers. The cavity presented in this paper has $L_{RT} = 6~\mathrm{km}$ and $L_{ex}=1~\mathrm{km}$ yielding $\tau_{max}= 6.66~\mu \mathrm{s}$, just exceeding the selected pulse duration of $\tau = 6~\mu \mathrm{s}$.

\paragraph{Control of frequency and phase}
Frequency and phase shifts arise during the sequence that require active control. There are two sources of frequency shift both caused by the Doppler effect: photon recoil and gravitational acceleration. Recoil shifts are caused by the absorption or emission of a photon, and affect the fast-arm of the interferometer during the main LMT sequence. Each recoil event causes a change in the velocity of the atoms, leading to a Doppler shift of $\hbar k^2/m=2\pi \cdot 9.4~\mathrm{kHz}$ per recoil. Gravity accelerates both arms of the interferometer downwards and causes an additional Doppler shift of $14~\mathrm{MHz} /\mathrm{s}$, or 281~Hz per round trip. We track these shifts by applying serrodyne modulation to each of the two circulating pulses. When a pulse travels through the Pockels cell a serrodyne shift is applied. High efficiency modulation is achieved by applying a single, linear ramp for the duration of the pulse~\cite{Tomita_2005}. This is a $\chi^{(2)}$ effect, so the phase shift $\phi$, is proportional to applied voltage:

\begin{equation}
    \phi = \pi \frac{V(t)}{V_\pi}= \pi \frac{V_0 + kt}{V_\pi}=  \phi_0 + \frac{\pi k t}{V_\pi}
    \label{eq:Serrodyne} 
\end{equation}

Frequency is defined as $d\phi/dt$ so a linear phase chirp produces a frequency shift. By adjusting the value of k the frequency can be shifted by up to $1/\tau~\mathrm{Hz}$ for a standard length Pockels cell, (ramping from $\phi = -\pi \to \pi$ over the pulse duration $\tau$ ) and up to $N/\tau~\mathrm{Hz}$ for an extended length $N \pi$-Pockels cell. The performance of this scheme is limited by the modulation efficiency and any losses associated with the modulator itself. Both circulating modes use the same polarization of light, so reflection losses can be reduced by using a Pockels cell with Brewster's angle cut crystal facets~\cite{immarco1971Brewster}.

Phase compensation can be achieved in two ways. DC voltages can be applied to the intracavity Pockels cell to ensure that the phase on the input mirrors is constant for every pulse in both circulating modes. The voltages required will differ every round trip, and be different for each circulating mode. Longer Pockels cells will increase flexibility to introduce a phase shift whilst leaving enough headroom for modulation. A simpler approach experimentally, is to adjust the phase of the input pulses on each round trip to match that of the circulating field. A Pockels cell or acousto-optic modulator on each of the input beams could achieve this in an agile and controllable way. Regardless of which phase compensation technique is adopted, the frequency of the input beams must also be tuned to track the circulating fields.

\paragraph{Laser system}
This scheme places demanding constraints on the input lasers, requiring narrow linewidth, rapid tunability, and high output power. It will require at least two separate lasers to inject light into both circulating modes of the cavity. The coherence length of the lasers must exceed the average distance traveled by a photon within the cavity. This ensures that successive input pulses continue to constructively interfere with the intracavity circulating pulse. With a cavity photon lifetime of $t_c = L\mathcal{F}/\pi c \simeq 30 ~\mathrm{ms}$, a laser linewidth of $\sim 1~\mathrm{Hz}$ will ensure the limit is comfortably met. The optical paths on the input to the cavity must also be phase stable at $\sim1~\mathrm{Hz}$ level. These performance levels are demanding, but regularly achieved in optical clock lasers operating on the 698~nm transition~\cite{Ludlow06SrHz,LIGOLinewidith2009}.

Clock lasers do not typically produce the output powers required for this scheme, so we propose the use of a clock laser as the master to injection-lock a high power slave~\cite{Cummings:02}. Commercially available high power Ti:sapphire lasers generate $\geq 5\mathrm{W}$, so the required output of $16~\mathrm{W}$ would require the coherent combination of four of these. The pulsed nature of the output may allow the use of a single Q-switched slave laser, reducing the system complexity. Injection locking also simplifies frequency and phase compensation which may be accomplished by appropriate modulation of the master laser, prior to injection seeding. This retains the stability properties of the master laser and allows modulation to occur at low power levels.

\bibliography{references}
\bibliographystyle{ScienceAdvances}

\noindent \textbf{Acknowledgments:} R.N. devised the circulating pulse scheme. S.L. derived the theoretical framework and code, and performed the calculations. S.H. and M.L. contributed to the LMT sequence design and Doppler shifts compensation techniques. K.B. and M.H. contributed throughout and supervised the research. All authors contributed to reviewing and assessing the results, and to the development and review of the manuscript. We thank Y.-H. Lien and J. Goldwin for helpful discussions.\\
\noindent \textbf{Funding:} This work was supported by the EPSRC under grant number EP/T001046/1. \\
\noindent \textbf{Data and materials availability:} All data needed to evaluate the conclusions in the paper are present in the paper and/or the Supplementary Materials. Additional data related to this paper may be requested from the authors. \\
 \noindent \textbf{Competing Interests:} The authors declare that they have no competing interests. \\
\end{document}